\documentclass[pr,nofootinbib,twocolumn,showpacs,aps]{revtex4-1}
\usepackage{bm}
\usepackage{graphicx}
\usepackage{amssymb}
\usepackage{amsmath}
\usepackage{eufrak}
\usepackage{color}
\usepackage[utf8]{inputenc}
\usepackage[unicode=true,colorlinks=true,urlcolor=blue,citecolor=blue]{hyperref}

\usepackage{pifont}
\usepackage{ulem}

\begin{document}

\title{Nonlinear Absorption and Photocurrent in Weyl Semimetals}

\author{N. V. Leppenen,
  E. L. Ivchenko,
  L. E. Golub}

\affiliation{Ioffe Institute, 194021 St.~Petersburg, Russia}

\begin{abstract}
Theory of light absorption and circular photocurrent in Weyl semimetals is developed for arbitrary large light intensities with account for both elastic and inelastic relaxation processes of Weyl fermions. The direct optical transition rate is shown to saturate at large intensity, and the saturation behaviour depends on the light polarization  and on the ratio of the elastic and inelastic relaxation times. The linear-circular dichroism in absorption is shown to exceed 10~\%  at intermediate light wave amplitudes and fast energy relaxation. At large intensity $I$, the light absorption coefficient drops as $1/\sqrt{I}$, and the circular photogalvanic current increases as $\sqrt{I}$.
\end{abstract}

\maketitle{}

\section{Introduction}

Weyl semimetals are extremely interesting systems representing a solid-state realization of three-dimensional fermions with linear energy spectrum. The specific feature of Weyl semimetals is the Circular Photo-Galvanic Effect (CPGE) -- generation of the photocurrent which inverses its direction at inversion of the  helicity of the absorbed light. The specifics is that the CPGE current density  has a ``quantized'' generation rate which, except for light intensity, is given by a combination of fundamental constants independent of both details of the real bandstructure of the system and the light frequency~\cite{Moore}. This quantization has been shown to be robust to electron-electron interaction which yields a very weak frequency dependence additionally suppressed by a factor equal to an inverse number of Weyl nodes~\cite{GolubIvch}. Recently this universal photocurrent has been detected in the Weyl semimetal RhSi with a good agreement in the value of the generation-rate quantization~\cite{RhSi}.
In the lower frequency range, where the indirect absorption takes place, the CPGE current density is determined by the light intensity, frequency and fundamental constants~\cite{GolubIvch}. Under unpolarized excitation, the photocurrent is generated in the presence of magnetic field, where it is enhanced if one of photocarriers is excited to or from the chiral magnetic subband in the quantizing magnetic field~\cite{GolubIvchSpivak_JETP_Lett}.
 
\begin{figure}[htb]%
\centering
\includegraphics*[width=.2\textwidth]{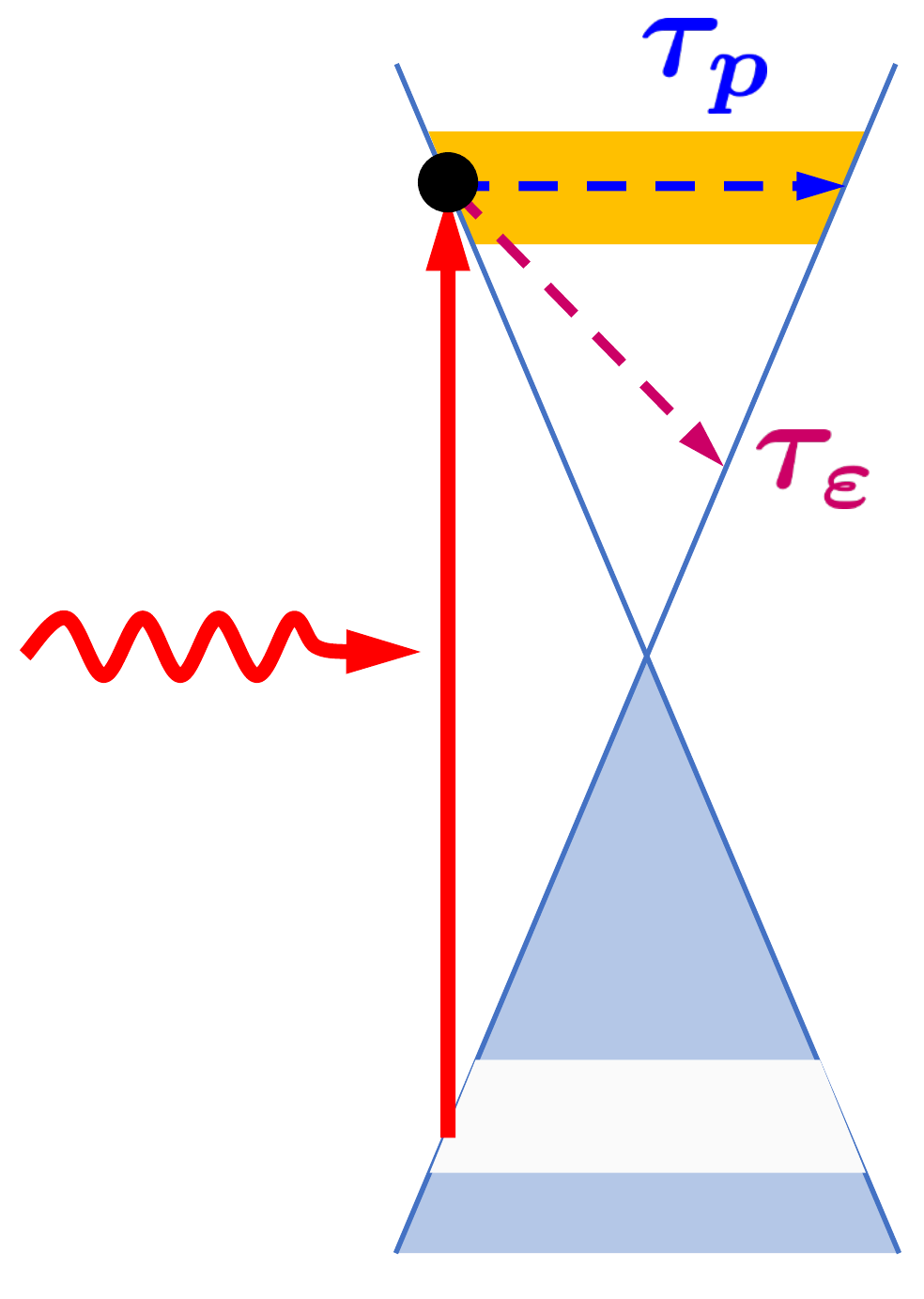}
\caption{%
Scheme of direct optical transitions. The photoexcited carriers experience the intraband momentum and energy relaxation processes with the relaxation times $\tau_p$ and $\tau_\varepsilon$, respectively.}
\label{fig_scheme}
\end{figure}

In the present work we develop a theory of light absorption  and CPGE in Weyl semimetals at arbitrarily large light intensity. In fact, this problem comes down to a consideration of an ensemble of two-level systems, the pairs of conduction and valence band states with the same wave vector ${\bm k}$. For many physical considerations of light-matter interaction it is sufficient to take only two energy eigenstates into account, namely, the states $|1\rangle$ and $|2\rangle$ with the energy difference $E_2 - E_1$ close to the incident light photon energy $\hbar \omega$. The pair states can be two levels in atoms \cite{Boyd,UFN2}, spin-split states in the magnetic resonance \cite{Abraham,magnetic,UFN}, tunneling two-level systems
in dielectric glasses~\cite{glass,Phillips}, the ground electron and hole quantum-confined levels in semiconductor quantum dots \cite{QD2008,Dnepr} etc. In such systems the saturation of resonant optical properties is controlled by two characteristic times, longitudinal and transverse relaxation times $T_1$ and $T_2$ respectively ($T_2 \leq 2T_1$). Another realization of two-level quantum-mechanical systems is a pair of electronic bands in a solid. For direct optical transitions the band states $|1, {\bm k}\rangle$ and $|2, {\bm k}\rangle$, for example the states in the heavy- ($hh$) and light-hole ($lh$) subbands of germanium, are characterized by a continuous wave vector ${\bm k}$ and thus form an ensemble of two-level systems \cite{Parshin_Shabaev_JETP_1987,Parshin}. In this case the saturation of resonant optical transitions is governed by the hole lifetimes $\tau_1$ and $\tau_2$ in the states $|hh, {\bm k}\rangle$ and $|lh, {\bm k}\rangle$ with a fixed value of the wave vector because any scattering process is assumed to take the hole out of resonance: due to the warping of the valence band even an elastic scattering removes a hole from the resonance transition region. The Weyl semimetals present a class of solids with an ensemble of two-level electronic states where elastic scattering leaves the resonantly excited electrons and holes in the resonant region. Indeed, even for the anisotropic effective Hamiltonian 
\begin{equation} \label{beta}
{\cal H} = \beta_{ij} \sigma_{i}k_{j}
\end{equation}
in the Weyl semimetal and anisotropic energy spectrum 
\begin{equation} \label{Lambda}
E_{c {\bm k} } = - E_{v {\bm k}} = \sqrt{\Lambda_{j'j} k_{j'}k_j}\:,
\end{equation}
the electron and hole energies are given by $\hbar \omega/2$ and, therefore, the elastic scattering process ${\bm k} \to {\bm k}'$ retains their energies refered to the Weyl node. Here $\sigma_i$ are the Pauli matrices, $\hat{\bm \Lambda}=\hat{\bm \beta}^T \hat{\bm \beta}$ is a symmetric $3 \times 3$ matrix, and $\hat{\bm \beta}^T$ is a transpose of the matrix $\hat{\bm \beta}$ \cite{JETP_2019}. As a result  the energy and momentum relaxation times, $\tau_{\varepsilon}$ and $\tau_p$, should influence differently the saturation behaviour: the energy relaxation removes the photocarries from the resonance regiont marked by yellow and white stripes in Fig.~\ref{fig_scheme}
and resists to the saturation whereas the elastic scattering causes the off-diagonal electron-hole decoherence but, on the other hand, conserves the carrier population in the resonant regions, the latter promoting the saturation. 

For simplicity, in the main part of the article, we assume the isotropic electron energy spectrum with $\hat{\bm \Lambda}$ being proportional to the identity matrix, i.e., $\Lambda_{j'j} = (\hbar v_0)^2 \delta_{j'j}$ in which case, instead of the general equation (\ref{Lambda}), one has
\begin{equation}
E_{ck} = -E_{vk} = \hbar v_0 k\:.
\end{equation}
Here $v_0>0$ is the Weyl fermion velocity and $k$ is the absolute value of the vector $\bm k$. At the final part we briefly present the result for an arbitrary matrix $\hat{\bm \Lambda}$. 
For the sake of definiteness, we consider the contribution of one Weyl cone with positive chirality.

\section{Distribution of photocarriers in nonlinear regime}

The rate of direct optical transition rate $W_{cv}$ between the valence ($v$) and conduction ($c$) bands is given by~\cite{Boyd}
\begin{equation}
\label{W_cv}
W_{cv}(\bm k) = G_{\bm k}(f_{v \bm k}-f_{c \bm k}),
\end{equation}
where  $f_{c,v}$ are the distribution functions in the bands, and the generation rate is
\begin{equation}
\label{G}
G_{\bm k} ={2|M_{cv}(\bm k)|^2/\tau \over (E_{ck}-E_{vk}-\hbar\omega)^2 + (\hbar/\tau)^2}.
\end{equation}
Here $\omega$ is the light frequency, $M_{cv}$ is the matrix element of the direct transition,  and $\tau$ is the relaxation time
\begin{equation}
\label{times}
{1\over \tau} = {1\over \tau_p} + {1\over \tau_\varepsilon}
\end{equation}
with $\tau_p$ and $\tau_\varepsilon$ being the momentum and energy relaxation times, respectively, which are equal for electrons and holes due to the electron-hole symmetry of the Weyl Hamiltonian~\eqref{beta}.
We assume that the energy relaxation process removes the carrier from the resonant region.
Note that we take into account only resonant contribution to the optical transition rate while the smooth contribution~\cite{Parshin_Shabaev_JETP_1987} is beyond the scope of the present paper.

For elliptically polarized excitation, the optical matrix element squared is given by
\begin{align}
\label{M_cv}
 & |M_{cv}(\bm k)|^2= \left({eE_0 v_0 \over \omega}\right)^2 \\ 
& \times\left[1 + \varkappa \cos{\theta_{\bm k}} - {\sin^2{\theta_{\bm k}}\over 2}(1+P_l\cos{2\varphi_{\bm k}})\right], \nonumber
\end{align}
where $\theta_{\bm k}$, $\varphi_{\bm k}$ are spherical angles of the wavevector $\bm k$ in the coordinate system with the $z$ axis along the light propagation direction and the $x$ axis is the polarization ellipse major axis, $E_0$ is  the light amplitude, and $P_l$ and $\varkappa$ are the light linear polarization degree and helicity, respectively. For fully polarized light, $P_l^2+\varkappa^2=1$.

In order to find the nonequilibrium distributions  in the conduction and valence  bands  we use the kinetic equation: 
\begin{equation}
\label{kin_eq}
{f_{c \bm k}-f_c^0\over \tau_\varepsilon} + {f_{c \bm k}-\left<{f_{ck}}\right>\over \tau_p} = W_{cv}(\bm k) .
\end{equation}
Here the angular brackets denote averaging over directions of $\bm k$ at a fixed absolute value $k$.
The distribution functions are given by
\begin{equation}
f_{i \bm k} = f_i^0 + \Delta f_{i \bm k},  \qquad \Delta f_{c \bm k}=-\Delta f_{v \bm k},
\end{equation}
where $i=c,v$, $f_i^0$ are equilibrium occupations (in the dark), and the last equality follows from the electron-hole symmetry.
Solution of the kinetic equation yields
\begin{align}
\label{dfck}
\Delta f_{c \bm k} & = {f_v^0-f_c^0\over 2}\\ 
&\times\left( {G_{\bm k}\over 1/2\tau + G_{\bm k}}{1\over 1+\Psi_k\tau_\varepsilon/\tau_p} + {\Psi_k\over \Psi_k + \tau_p/\tau_\varepsilon} \right), \nonumber
\end{align}
where
\begin{equation}
\label{Psi}
\Psi_k = \left< {G_{\bm k} \over 1/2\tau + G_{\bm k}}\right>.
\end{equation}
At large light intensity when $G_{\bm k} \gg 1/\tau$, the nonequilibrium correction $\Delta f_{c \bm k}  \approx (f_v^0-f_c^0)/2$, so the occupations in the bands are equal to each other: $f_{c \bm k}=f_{v \bm k}=(f_v^0+f_c^0)/2$, and the direct optical transition rate~\eqref{W_cv} vanishes.

\section{Nonlinear absorption}

The absorption coefficient $\alpha$ is related with the direct interband transition rate and the intensity
$I=cnE_0^2/(2\pi)$, where $n$ is the refractive index at the frequency $\omega$,
as follows:
\begin{equation}
\label{alpha_general}
{\alpha I\over \hbar\omega} = \sum_{\bm k} W_{cv}(\bm k) = \sum_{\bm k} G_{\bm k} (f_v^0-f_c^0-2\Delta f_{c \bm k}).
\end{equation}
Substitution of the nonequilibrium part $\Delta f_{c \bm k}$ from Eq.~\eqref{dfck} yields
\begin{equation}
\label{alpha1}
{\alpha I\over \hbar\omega} 
= {\cal F} \sum_{\bm k} {\Psi_k/2\tau\over 1+\Psi_k\tau_\varepsilon/\tau_p},
\end{equation}
where ${\cal F}=f_0(-\hbar\omega/2)-f_0(\hbar\omega/2)$ with $f_0$ being the Fermi-Dirac distribution function. Hereafter we use the fact that $\Psi_k$ has a sharp maximum at $k=\omega/(2v_0)$ provided $\omega\tau \gg 1$.
Extending integration over the variable $k-\omega/(2v_0)$ to the whole real axis, we obtain
\begin{equation}
\label{alpha}
\alpha = \alpha_0{6\over \pi }\int\limits_0^{\cal E} ds {\Psi(s)\over s^2\sqrt{{\cal E}^2-s^2}[ 1+\Psi(s)\tau_\varepsilon/\tau_p]} .
\end{equation}
Here we introduce the dimensionless electric field amplitude
\begin{equation}
\label{E_cal}
\mathcal{E}={2\sqrt{2}eE_0v_0\tau/\hbar \omega},
\end{equation}
and the variable $s=\mathcal{E}[1+(\omega-2v_0k)^2\tau^2]^{-1/2}$.
The absorption coefficient at low intensity, $\alpha_0$, is given by
\begin{equation}
\label{alpha_0}
\alpha_0 = {\cal F}{e^2\omega \over 6 \hbar c v_0}.
\end{equation}
It is related to the optical conductivity 
$\sigma$ via $\alpha_0 I = 2\sigma E_0^2$. This yields at ${\cal F}=1$ the value $\sigma = e^2\omega/(12 h v_0)$ 
coinciding with the contribution of one Weyl cone, see e.g. Ref.~\cite{cond}.
It follows from Eq.~\eqref{alpha} that the absorption coefficient drops as $1/{\cal E} \propto 1/\sqrt{I}$ at ${\cal E} \to \infty$.

\begin{figure}[htb]%
\centering
\includegraphics*[width=.38\textwidth]{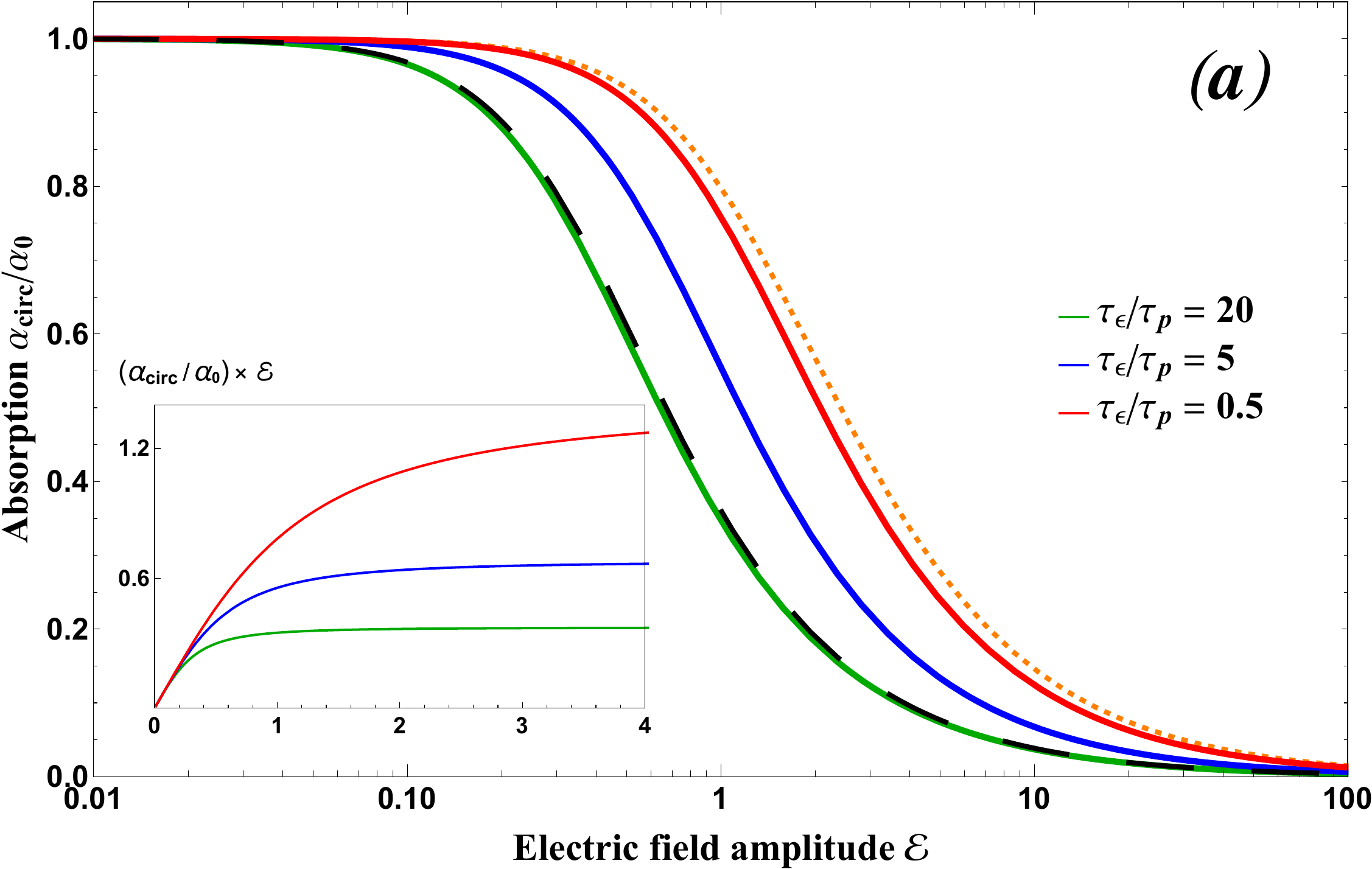}\\
\includegraphics*[width=.38\textwidth]{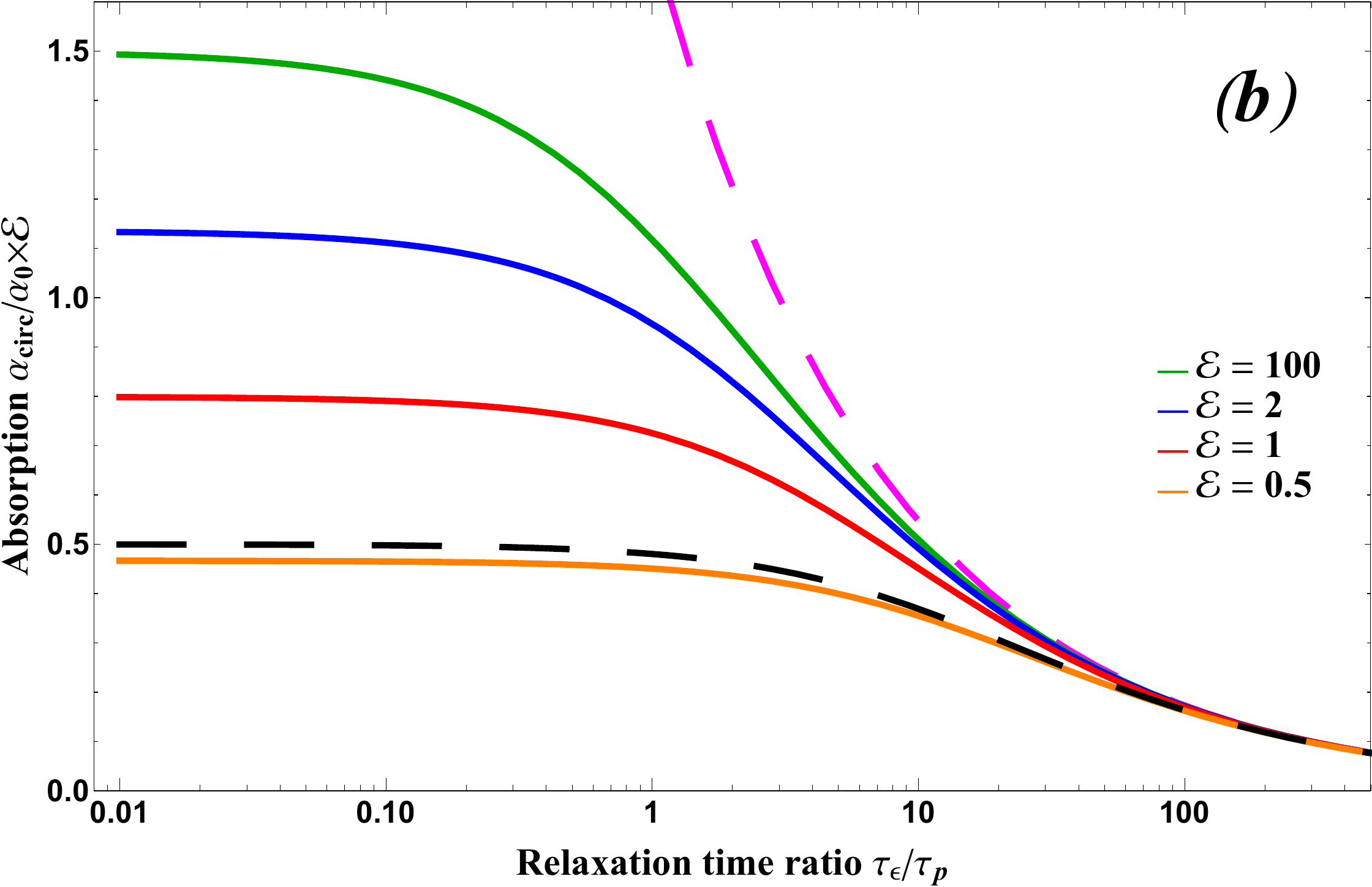}
\caption{%
Absorption coefficient at right- or left-handed circular polarization.
(a)~The absorption coefficient  
as a function of the dimensionless light electric field ${\cal E}$ at various ratios between the energy and momentum relaxation times. Black dashed curve is the approximation~\eqref{alpha_appr}, and the orange dotted curve is  an analytical result~\eqref{alpha_an}. Inset: The electric-field dependence of $(\alpha_\text{circ}/\alpha_0)  {\cal E}$ demonstrating the $1/{\cal E}$ law for bleaching.
(b)~The dependence of the absorption coefficient  on the ratio of the energy and momentum relaxation times at various values of the light amplitude ${\cal E}$. Black and magenta dashed lines are the approximation~\eqref{alpha_appr} calculated at ${\cal E}=0.5$ and ${\cal E}=100$, respectively.}
\label{alpha_vs_E}
\end{figure}

For circularly polarized light we obtain from Eqs.~\eqref{G}, \eqref{M_cv} and~\eqref{Psi} that $\Psi_\text{circ}$ 
is given by
\begin{equation}
 \Psi_\text{circ}(s)= 1 - {\arctan{s}\over s} .
\end{equation}
Substitution to Eq.~\eqref{alpha} yields the dependence of the absorption coefficient on the electric field amplitude and the relaxation times. The dependence $\alpha({\cal E})$ is presented in Fig.~\ref{alpha_vs_E}~(a) for various ratios $\tau_\varepsilon/\tau_p$. Inset to Fig.~\ref{alpha_vs_E}~(a) demonstrates the law $\alpha \propto 1/{\cal E}$ at large light intensity. Figure~\ref{alpha_vs_E}~(b) shows the dependence of $\alpha$ on $\tau_\varepsilon/\tau_p$ at fixed values of ${\cal E}$.

In the limiting cases of slow and fast energy relaxation we obtain analytical results.
Hereafter we assume that the total relaxation rate $1/\tau$ is fixed, while the relation between the elastic and inelastic relaxation times, $\tau_p$ and $\tau_\varepsilon$, can be arbitrary.
At $\tau_\varepsilon/\tau_p \to 0$ we have from Eq.~\eqref{alpha}
\begin{equation}
\label{alpha_an}
{\alpha_\text{circ} \over \alpha_0} = {3\over2}\frac{\mathcal{E} \sqrt{\mathcal{E}^2+1}-\text{arcsinh}\:{\mathcal{E}} }{\mathcal{E}^3}.
\end{equation}
In the opposite limit
$\tau_\varepsilon/\tau_p \to \infty$, the main contribution to the integral comes from  $s \ll 1$ 
due to the factor $s^{-2}$.
Since $\Psi_\text{circ}(s\ll 1)\approx s^2/3$, we obtain 
\begin{equation}
\label{alpha_appr}
{\alpha_\text{circ} \over \alpha_0} 
\approx {1\over \sqrt{1+{\cal E}^{2}\tau_{\varepsilon}/3\tau_{p}}}.
\end{equation}
Figure~\ref{alpha_vs_E}~(a) shows that these expressions describe the exact dependence $\alpha_\text{circ}({\cal E})$ with a high accuracy at $\tau_\varepsilon/\tau_p \leq 0.5$ and $\tau_\varepsilon/\tau_p \geq 20$, respectively. 
The dashed lines in Fig.~\ref{alpha_vs_E}~(b) demonstrate validity of approximation~\eqref{alpha_appr} for description of the dependence $\alpha_\text{circ}(\tau_\varepsilon/\tau_p)$.

At low intensity, the  absorption coefficient equals to $\alpha_0$ at any polarization. In the nonlinear in intensity regime, by contrast, the linear-circular dichroism takes place. 
For linearly polarized light we have from Eqs.~\eqref{G}, \eqref{M_cv} and~\eqref{Psi}
\begin{equation}
\Psi_\text{lin}(s) = 1+ {\ln{2}-2\ln{\left(s+\sqrt{s^2+2}\right)} \over s\sqrt{s^2+2}},
\end{equation}
and from Eq.~\eqref{alpha} we obtain the dependence $\alpha_\text{lin}({\cal E},\tau_\varepsilon/\tau_p)$.

\begin{figure}[htb]%
\centering
\includegraphics*[width=.38\textwidth]{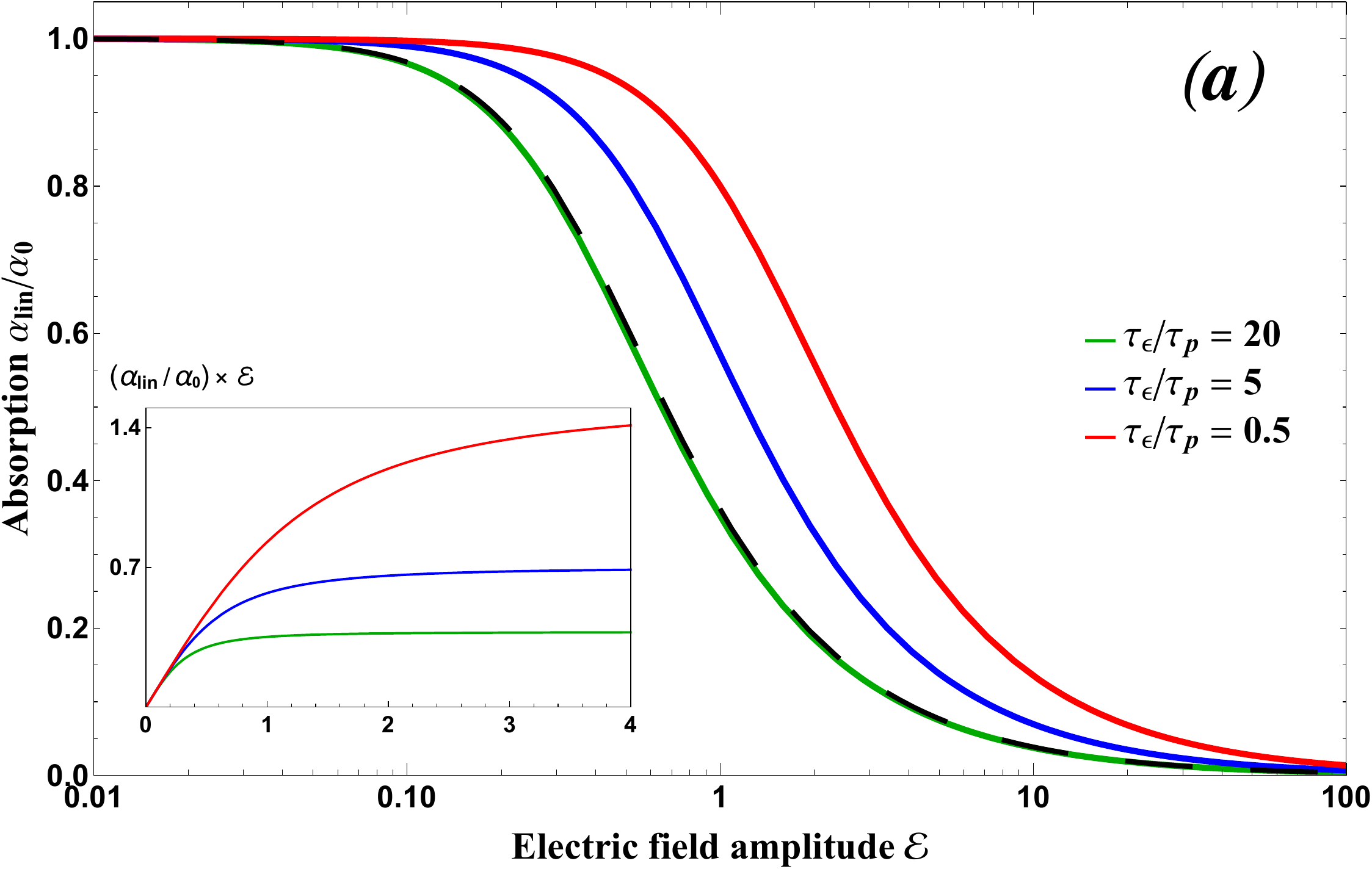}\\
\includegraphics*[width=.38\textwidth]{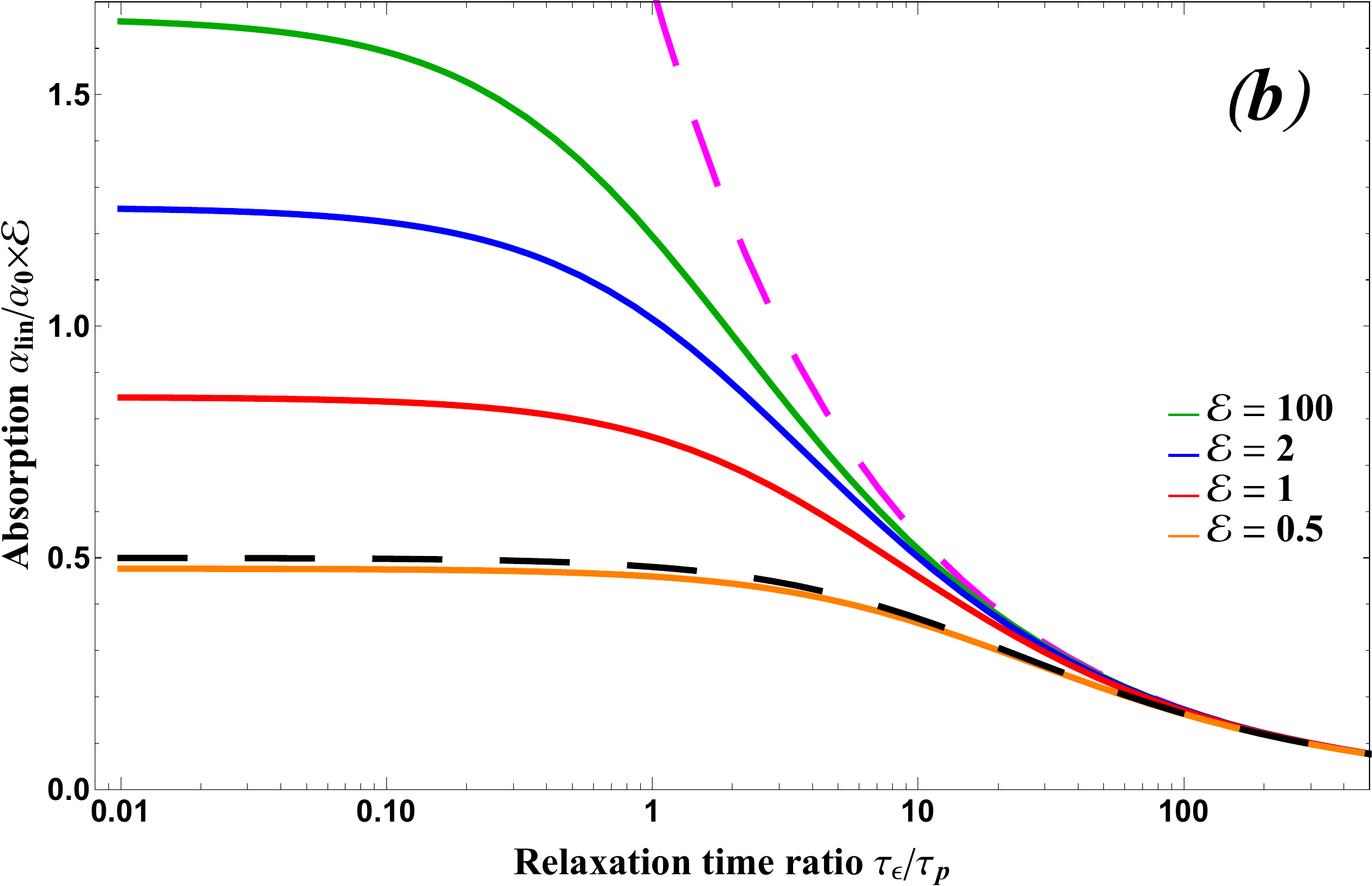}
\caption{%
Absorption coefficient at linear polarization.
(a)~The absorption coefficient  
as a function of the dimensionless electric field of light ${\cal E}$ at various ratios between the energy and momentum relaxation times. 
Inset: The electric-field dependence of $(\alpha_\text{lin}/\alpha_0)  {\cal E}$ demonstrating the $1/{\cal E}$ law for bleaching. 
(b)~The dependence of the absorption coefficient  on the ratio of the energy and momentum relaxation times at various values of ${\cal E}$. Black and magenta dashed lines are the approximation~\eqref{alpha_appr} calculated at ${\cal E}=0.5$ and ${\cal E}=100$, respectively.
}
\label{alpha_vs_E_lin_pol}
\end{figure}

The dependence of the absorption coefficient at linear polarization on the electric field amplitude is shown in Fig.~\ref{alpha_vs_E_lin_pol}~(a). At large $\tau_\varepsilon/\tau_p$, the absorption coefficient $\alpha_\text{lin}$ has the same asymptotics~\eqref{alpha_appr} as $\alpha_\text{circ}$.
Figure~\ref{alpha_vs_E_lin_pol}~(b) demonstrates the dependence $\alpha_\text{lin}$ on $\tau_\varepsilon/\tau_p$. Bleaching at slow energy relaxation is clearly seen from decrease of the absorption coefficient at $\tau_\varepsilon \gg \tau_p$.

\begin{figure}[htb]%
\centering
\includegraphics*[width=.38\textwidth]{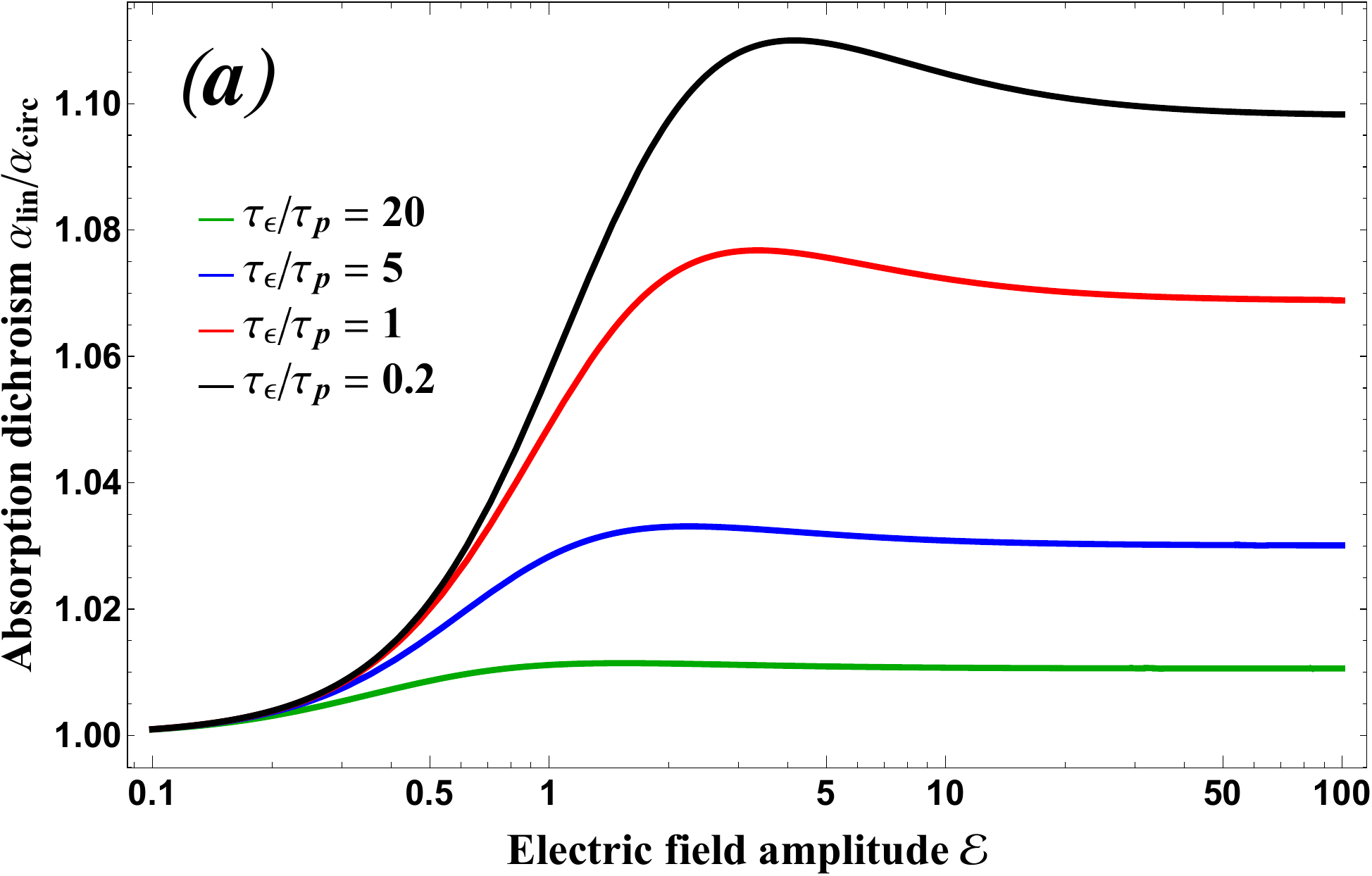}\\
\includegraphics*[width=.38\textwidth]{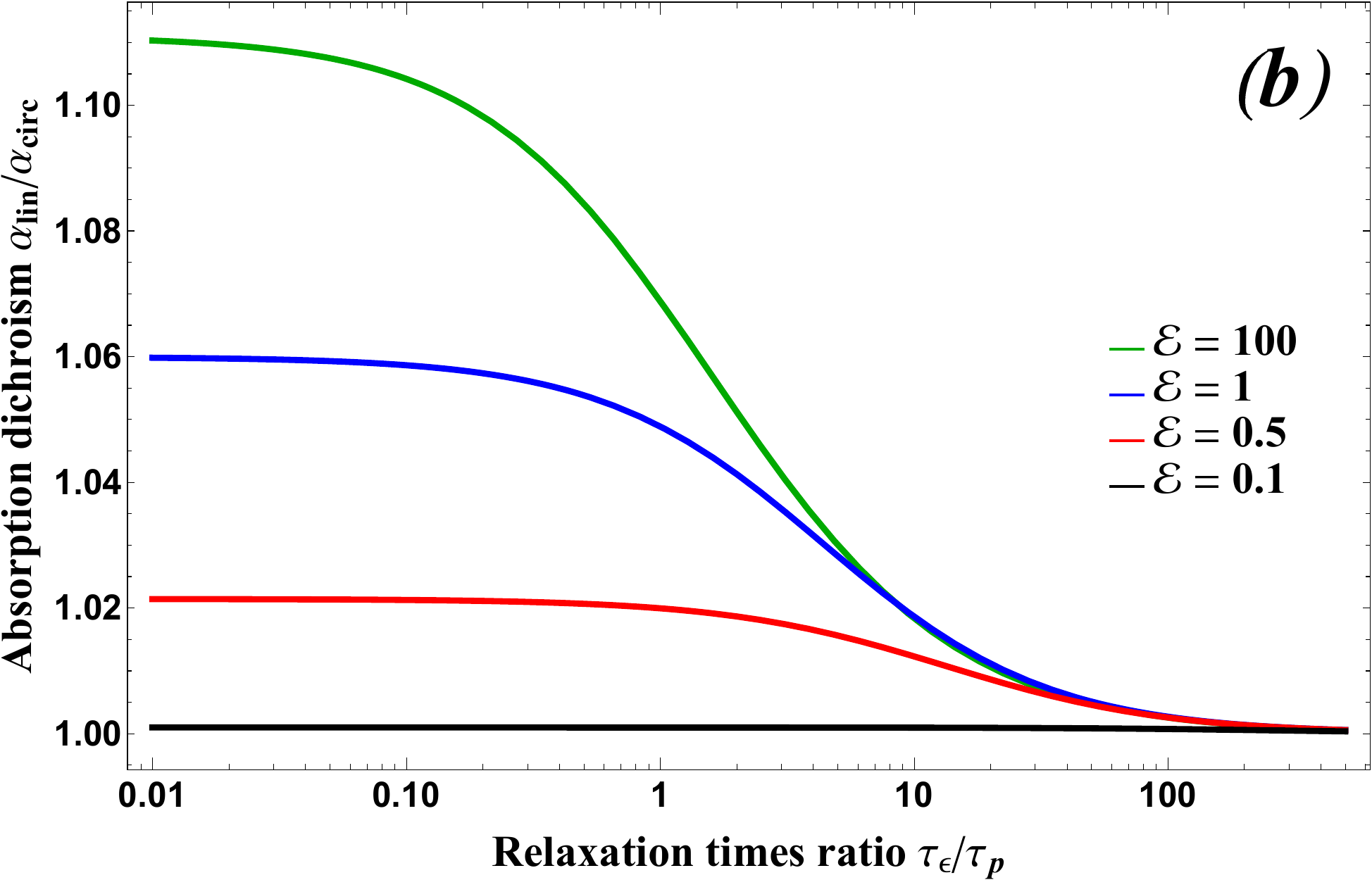}
\caption{%
Linear-circular dichroism in absorption: the ratio $\alpha_\text{lin}/\alpha_\text{circ}$ as a function of the dimensionless light wave amplitude ${\cal E}$ (a) and $\tau_\varepsilon/\tau_p$ (b).}
\label{dichroism}
\end{figure}

In Fig.~\ref{dichroism}, the ratio $\alpha_\text{lin}/\alpha_\text{circ}$ is plotted as a function of ${\cal E}$ and $\tau_\varepsilon/\tau_p$. This dependence shows that the degree of the linear-circular  dichroism in Weyl semimetals is about 10~\% at small $\tau_\varepsilon/\tau_p$ and large ${\cal E}$.

\section{Nonlinear Circular photocurrent}

The photocurrent density is given by
\begin{equation}
\bm j = 2e \sum_{\bm k} v_0{\bm k\over k} \Delta f_{c \bm k},
\end{equation}
where the factor 2 accounts for the contribution of photoholes.
From Eq.~\eqref{dfck} we obtain
\begin{equation}
\label{j_nl}
\bm j = \bm j_1 {12\over \pi} \int\limits_{0}^{\cal E} ds {\Phi(s)\over s^2\sqrt{{\cal E}^2-s^2}[ 1+\Psi_\text{circ}(s)\tau_\varepsilon/\tau_p]}.
\end{equation}
Here $\bm j_1$ is the ``quantized'' CPGE current linear in the light intensity~\cite{Moore}
\begin{equation}
\label{j1}
\bm j_1 = \bm \varkappa {\pi e^3\over 3h^2}|E_0|^2 \tau {\cal F},
\end{equation}
where  $\bm \varkappa$ is the photon vector helicity, and the function $\Phi$ is introduced according to
\begin{equation}
\bm \varkappa \Phi(s) =\left< {\bm k\over k}{G_{\bm k}  \over 1/2\tau + G_{\bm k}}\right> .
\end{equation}
From Eqs.~\eqref{G} and~\eqref{M_cv} we obtain
\begin{equation}
\Phi(s) = {\arctan{s}\over s} - {\ln{(1+s^2)}\over s^2}.
\end{equation}

\begin{figure}[htb]%
\centering
\includegraphics*[width=.38\textwidth]{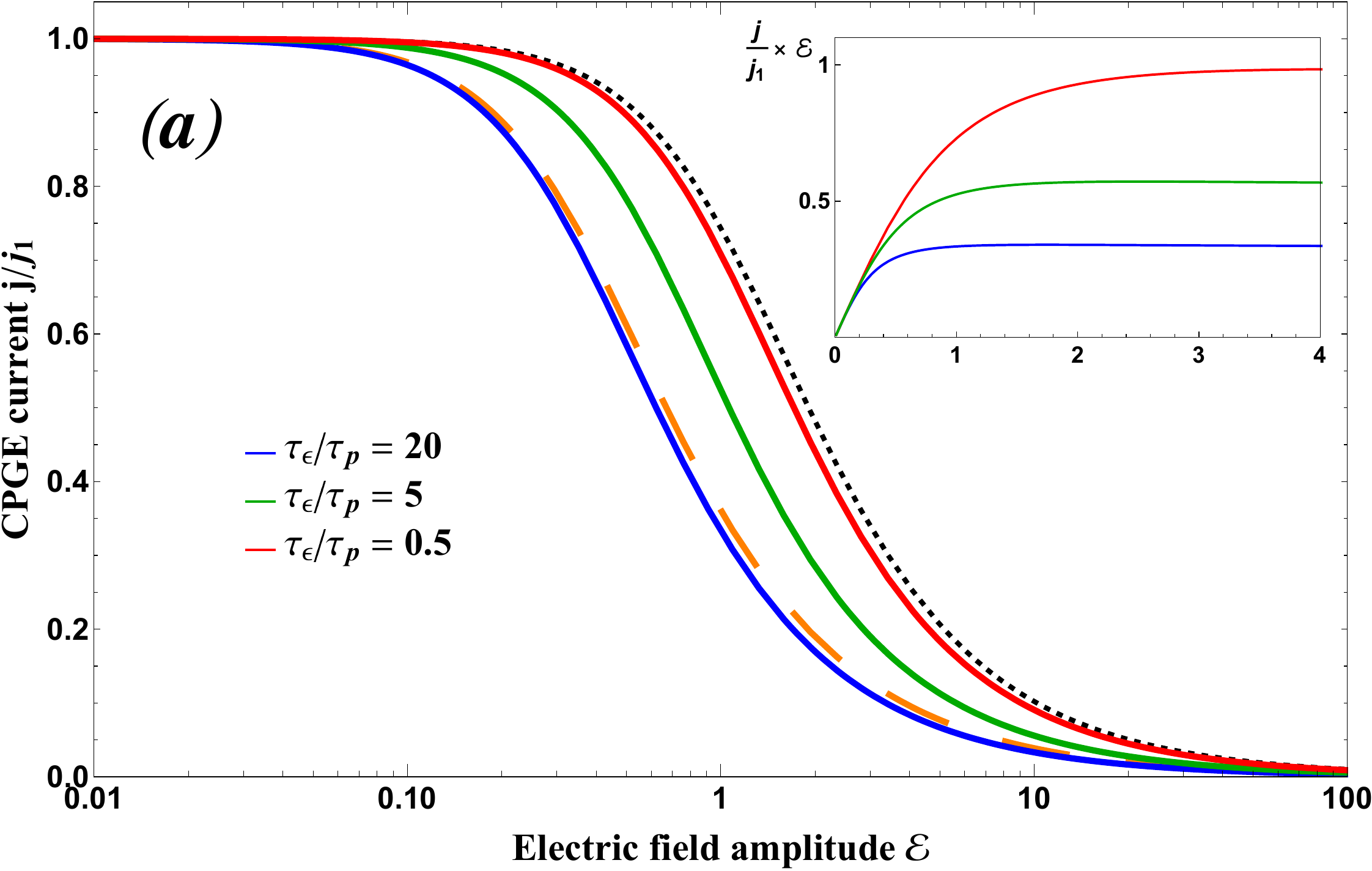}\\
\includegraphics*[width=.38\textwidth]{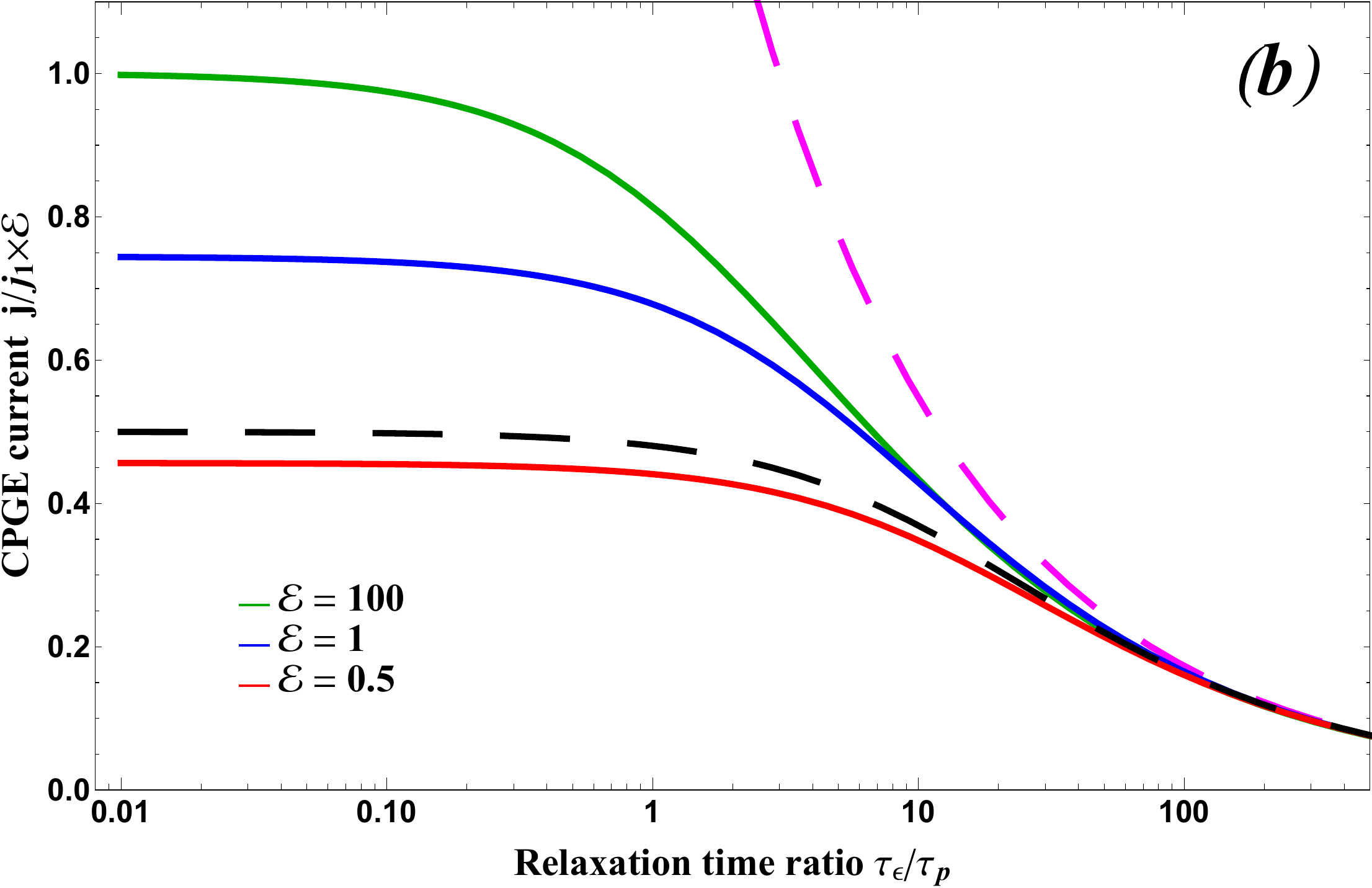}
\caption{%
(a)~CPGE current dependence on the dimensionless electric field of light ${\cal E}$ at various ratios between the energy and momentum relaxation times. The CPGE current density is normalized to its value in the linear regime. 
Black dotted curve is an analytical result~\eqref{current_an} and the orange dashed curve is the approximation~\eqref{current_appr}. Inset: The electric-field dependence of $(j/j_1) {\cal E}$ demonstrating the $j \propto {\cal E}$ law at high light intensity.
(b)~The dependence of CPGE current on the ratio of the energy and momentum relaxation times at various light amplitudes ${\cal E}$. Black and magenta dashed lines are the approximation~\eqref{current_appr} calculated at ${\cal E}=0.5$ and ${\cal E}=100$, respectively.}
\label{j_vs_E}
\end{figure}

The electric-field dependence of the CPGE current is shown in Fig.~\ref{j_vs_E}. Similarly to the absorption coefficient, the ratio $j/j_1$  drops as $1/{\cal E}$ at high intensity.
At fast energy relaxation ($\tau_\varepsilon/\tau_p \to 0$) we have
\begin{equation}
\label{current_an}
{j\over j_1} = \frac{\sqrt{{\cal E} ^2+1} {\cal E} ^2-8
   \sqrt{{\cal E} ^2+1}+3 {\cal E} \text{arcsinh}\:{\cal E}+8}{{\cal E}^2}.
\end{equation}
 In the opposite limit with $\tau_{\varepsilon}\gg \tau_{p}$, the noticeable contribution to the integral~\eqref{j_nl} comes from ${\Psi_\text{circ}(s) \sim {\tau_{p}/ \tau_{\varepsilon}}\ll 1}$. Since $\Psi_\text{circ}(s)$ increases from 0 to 1 at $s\in(0,\infty)$, only small $s$ contribute to the current, and we can replace $\Psi_\text{circ}(s)$ and $\Phi(s)$ with their $s \rightarrow 0$ asymptotes $\Psi_\text{circ}(s) = {s^{2} /3}$ and $\Phi(s) = {s^{2}/6}$. Then
we obtain the asymptotics coinciding with that in Eq.~\eqref{alpha_appr}:
\begin{equation}
\label{current_appr}
{j\over j_1} \approx{1\over \sqrt{1+{\cal E}^{2}\tau_{\varepsilon}/3\tau_{p}}}.
\end{equation}
Figure~\ref{j_vs_E}~(a) shows that these limiting functions describe very well the electric-field dependencies of the CPGE current.
The dependence of the CPGE current on the ratio $\tau_{\varepsilon}/\tau_p$ is presented Fig.~\ref{j_vs_E}~(b). This plot demonstrates that the approximation~\eqref{current_appr} describes the photocurrent at  $\tau_{\varepsilon}/\tau_p \geq 50$.

\section{Discussion}

For arbitrary matrix $\hat{\bm \beta}$ in Eq.~\eqref{beta}, we have the following modifications of the results. Instead of Eq.~\eqref{M_cv} we obtain the squared matrix element of the direct optical transition in the following form:
\begin{align}
\label{M_cv_beta}
&|M_{cv}(\bm k)|^2 =  \left({eE_0 \over \hbar \omega}\right)^2\\ 
& \times \left[ e_ie_j^*\left(\Lambda_{ij}-\beta_{li}\beta_{nj}{Q_lQ_n\over Q^2}\right) + \varkappa_l \sqrt{\Delta_\Lambda}\beta_{ln}^{-1} {Q_n\over Q} \right], \nonumber
\end{align}
where the vector $\bm Q =\hat{\bm \beta}\bm k$ is introduced, and $\Delta_\Lambda=\text{det}{\hat{\bm \Lambda}}$. The absolute value of the vector $\bm Q$ is determined from the energy conservation as $Q=\hbar\omega/2$.
For the linear absorption coefficient we have instead of Eq.~\eqref{alpha_0} (we assume ${\cal F}=1$)
\begin{equation}
	\alpha_0 = {\omega e^{2}\over 6 c\sqrt{\Delta_{\Lambda}}}  \Lambda_{ij} e_{i}e_{j}^*. 
\end{equation}

We consider the limit of slow energy relaxation when $\tau_\varepsilon \gg \tau_p$ and derive the results in the second order in the light intensity.
In this limit, the main correction to the conduction-band distribution function is $\Delta f_{c\bm k}=-\tau_\varepsilon \left< G_{\bm k}\right>$, where averaging is performed at $Q=\hbar\omega/2$.
Therefore both the absorption coefficient and the CPGE current are calculated by the same expressions as in the linear in intensity regime but with the generation rate $G_{\bm k}(1-2\tau_\varepsilon\left< G_{\bm k}\right>)$. As a result, we obtain the first intensity-dependent correction to the absorption coefficient $\alpha_2$ in the form
\begin{equation}
	\alpha_2 
	=  -{2\tau_{\varepsilon}\tau_{p}e^{4}E_{0}^{2}\over 9 \omega c \hbar^{4}\sqrt{\Delta_{\Lambda}}} (\Lambda_{ij} e_{i}e_{j}^*)^{2},
\end{equation}
and for correction to the CPGE current:
\begin{equation}
\bm j_2= -\bm j_1 {4\tau_{p}\tau_{\varepsilon}\over3\hbar^{2}} \left( {e E_{0}\over \hbar \omega} \right)^{2}\Lambda_{ij} e_{i}e_{j}^*,
\end{equation}
where $\bm j_1$ is the universal current density which is given by Eq.~\eqref{j1}  at any tensor $\hat{\bm \beta}$.
At ${\Lambda_{ij}=(\hbar v_0)^2\delta_{ij}}$, these results coincide with the 
$\mathcal{E}^2$-contributions obtained from expansion of Eqs.~\eqref{alpha_appr} and~\eqref{current_appr} at $\mathcal{E} \to 0$: $\alpha_2/\alpha_0={j_2/j_1=-\mathcal{E}^2\tau_{\varepsilon}/(6\tau_p)}$.

The spin-independent $\bm k$-linear terms are present in real Weyl semimetals resulting in tilt of the electron dispersion. With account for tilt, elastic scattering processes remove photocarriers  off the resonance. Therefore elastic scattering affects the electron and hole distribution analogously to the energy relaxation processes considered above. As a result, the kinetic equation has the form of Eq.~\eqref{kin_eq} with the following modifications: the second term in the left-hand side is absent, and the total departure time $\tau$ should be taken instead of $\tau_\varepsilon$ in the first term. So, at large tilt,  the nonlinear absorption and CPGE current are described by the above developed theory at $\tau_\varepsilon/\tau_p=0$, but still $1/\tau=1/\tau_p+1/\tau_\varepsilon$. In particular, the absorption coefficient at circular polarization and the CPGE current are given by Eqs.~\eqref{alpha_an} and~\eqref{current_an}, respectively.

\section{Conclusion}

The developed theory demonstrates that light absorption in Weyl semimetals bleaches at high light intensities. The absorption coefficient in both linear and circular polarizations drops as $1/\sqrt{I}$, while the linear-circular dichroism takes place at intermediate intensities. The CPGE current raises as~$\propto \sqrt{I}$  at high intensity. Both the absorption coefficient and the CPGE current are sensitive to the ratio of the electron energy and momentum relaxation times.

\acknowledgments 
Financial support of the Russian Science Foundation (Project No. 17-12-01265) is acknowledged.  L.~E.~G. also thanks the Foundation for advancement of theoretical physics and mathematics ``BASIS''.

\end{document}